\documentstyle[aps,prl]{revtex}
\input{psfig.tex}
\begin{document}
\draft 
\twocolumn[\hsize\textwidth\columnwidth\hsize\csname @twocolumnfalse\endcsname
\title{Signature of Gravity Waves in Polarization of the Microwave
Background}

\author{Uro\v s Seljak\footnote{useljak@cfa.harvard.edu}}
\address{Harvard-Smithsonian Center for Astrophysics,
60 Garden Street, Cambridge, Massachusetts~~02138}
\author{Matias Zaldarriaga\footnote{matiasz@arcturus.mit.edu}}
\address{Department of Physics, MIT, Cambridge, Massachusetts~~02139}
\maketitle

\begin{abstract}
Using spin-weighted decomposition of polarization in 
the Cosmic Microwave Background (CMB)
we show that a
particular combination of Stokes $Q$ and $U$ parameters
vanishes for primordial fluctuations 
generated by scalar modes, but does not for those generated 
by primordial gravity waves.
Because of this gravity wave detection 
is not limited by cosmic variance as in the case of temperature
fluctuations.  
We present the exact expressions for various polarization power spectra, 
which are valid on any scale. 
Numerical evaluation
in inflation-based models shows that the expected signal 
is of the order of 0.5 $\mu K$, which could be directly tested in 
future CMB experiments.
\end{abstract}

]
\def\edth{\;\raise1.5pt\hbox{$'$}\hskip-6pt\partial\;}
\def\baredth{\;\overline{\raise1.5pt\hbox{$'$}\hskip-6pt
\partial}\;}
\def\bi#1{\bbox{#1}}
\def\gsim{\raise2.90pt\hbox{$\scriptstyle
>$} \hspace{-6.4pt}
\lower.5pt\hbox{$\scriptscriptstyle
\sim$}\; }
\def\lsim{\raise2.90pt\hbox{$\scriptstyle
<$} \hspace{-6pt}\lower.5pt\hbox{$\scriptscriptstyle\sim$}\; }

It is now well established that temperature 
anisotropies in CMB offer one of the best probes
of early universe, which could potentially lead to a precise
determination of a large number of cosmological parameters 
\cite{jungman,zss}. 
The main advantage of CMB versus more local probes of large-scale 
structure is that the fluctuations were created at an epoch when the
universe was still in a linear regime. While this fact has long been 
emphasized for temperature anisotropies, the same holds also
for polarization in CMB and as such it offers the same advantages
as the temperature anisotropies in the determination of cosmological 
parameters. The main limitation of polarization is that it is 
predicted to be
small: theoretical calculations show that CMB will be polarized 
at 5-10\% level on small angular scales and much less than that
on large angular scales \cite{pol,uros}. However, future
CMB missions (MAP, Planck) will be so sensitive that even such low signals
will be measurable.
Even if polarization by itself cannot compete with the
temperature anisotropies, a combination of the two could result in a 
much more accurate determination of certain cosmological parameters,
in particular those that are limited by a finite number of multipoles in 
the sky (i.e. cosmic variance).

Primordial gravity waves produce fluctuations in the tensor 
component of the metric, which could result in a significant 
contribution to the CMB anisotropies on large angular scales. 
Unfortunately, the presence of
scalar modes prevents one from clearly separating one contribution 
from another. If there are only a finite number of multipoles where 
tensor contribution is significant then there is a limit in amplitude 
beyond which 
tensors cannot be distingushed from random fluctuations.  
In a noise free experiment
the tensor to scalar ratio $T/S$ needs to be larger than 
0.15 to be measurable
in temperature maps \cite{knox}. 
Independent determination of
the tensor spectral slope $n_T$ is even less accurate and a rejection
of the consistency relation in inflationary models 
$T/S=-7n_T$ is only possible if $|n_T|\gg (T/S)/7$ \cite{knox,dodelson}.
Polarization produced by tensor
modes has also been studied \cite{tenspol}, 
but only in the small scale limit. 
In previous work correlations
between Stokes parameters $Q$ and $U$ have been used. 
These two variables are not the most suitable for the
analysis as they depend on the orientation of coordinate system.
It was recently shown \cite{uros} that in Fourier space $Q$ and $U$ 
can be decomposed in two components, which 
do not depend on orientation. Moreover,
scalar modes only contribute to one of the two, 
leaving the other as a probe of gravity waves. 
These arguments have been made in the small angle approximation. 
In this Letter we 
remove this limitation by presenting a full spherical analysis of
polarization using Newman-Penrose spin-s spherical harmonic decomposition. 
An alternative decomposition in terms of tensor harmonics has been 
presented recently by \cite{kks}. 
We show that there is  
a particular combination of Stokes parameters that vanishes in the case
of scalar modes, which can thus be used as a probe of gravity waves. 
We present the expression for the power spectrum of  
various polarization components
using the integral solution \cite{sz} and evaluate it
numerically for a variety of cosmological models. We also discuss
the sensitivity needed to detect this signal and compare it to the
expected sensitivities of future CMB satellites.

Linear polarization 
is a symmetric and traceless 2x2 tensor \cite{circular}
that requires 2 parameters to fully describe it: 
$Q$ and $U$ Stokes parameters. 
%In a local coordinate system $Q$ is positive if temperature anisotropy is larger
%along the $x$ relative to the $y$ axis, while $U$ parameter is positive
%if anisotropy is larger along the upper right diagonal relative to the 
%upper left diagonal.  
These parameters depend on the 
orientation of the coordinate system on the sky. It is convenient 
to use $Q+iU$ and $Q-iU$ 
as the two independent combinations, which
transform under right-handed 
rotation by an angle $\phi$ as $(Q+iU)'=e^{-2i\phi}(Q+iU)$ and
$(Q-iU)'=e^{2i\phi}(Q-iU)$. These two quantities therefore 
have spin-weights $2$ and $-2$ respectively and can be decomposed
into spin $\pm 2$ spherical harmonics ${}_{\pm 2}Y_{lm}$ (for a 
discussion of spin-weighted harmonics see 
\cite{goldberg}) 
\begin{eqnarray}
(Q+iU)(\hat{\bi{n}})&=&\sum_{lm} 
a_{2,lm}{}_2Y_{lm}(\hat{\bi{n}}) \nonumber \\
(Q-iU)(\hat{\bi{n}})&=&\sum_{lm}
a_{-2,lm} {}_{-2}Y_{lm}(\hat{\bi{n}}).
\end{eqnarray}
%on the sky.
Spin $s$ spherical harmonics form a complete orthonormal system 
for each value of $s$. 
An important property of spin-weighted basis is that there exists
spin raising and lowering operators $\edth$ and $\baredth$ 
(see \cite{goldberg} for their explicit form).
By acting twice with a spin lowering and raising 
operator on $(Q+iU)$ and $(Q-iU)$ respectively 
one obtains quantities of spin 0,
which are {\it rotationally invariant}. These quantities can be treated
like the temperature and no ambiguities connected with the orientation of 
coordinate system on the sky will arise. Conversely, by acting 
with spin lowering and raising operators on usual harmonics spin $s$ 
harmonics 
can be written explicitly in terms 
of derivatives of the usual spherical harmonics \cite{goldberg}.
Their action on ${}_{\pm 2}Y_{lm}$
leads to
\begin{eqnarray}
\baredth^2(Q+iU)(\hat{\bi{n}})&=&
\sum_{lm} 
\left({[l+2]! \over [l-2]!}\right)^{1/2}
a_{2,lm}Y_{lm}(\hat{\bi{n}})
\nonumber \\
\edth^2(Q-iU)(\hat{\bi{n}})&=&\sum_{lm} 
\left({[l+2]! \over [l-2]!}\right)^{1/2}
a_{-2,lm}Y_{lm}(\hat{\bi{n}}).
\end{eqnarray}
With these definitions the expressions for the expansion coefficients 
of the two polarization variables become
\begin{eqnarray}
a_{2,lm}&=&\left({[l-2]! \over [l+2]!}\right)^{1/2}
\int d\Omega\; Y_{lm}^{*}(\hat{\bi{n}}) \baredth^2(Q+iU)(\hat{\bi{n}})
\nonumber \\
a_{-2,lm}&=&\left({[l-2]! \over [l+2]!}\right)^{1/2}
\int d\Omega\; Y_{lm}^{*}(\hat{\bi{n}})\edth^2(Q-iU)(\hat{\bi{n}}). 
\label{alm}
\end{eqnarray}

To obtain the expression for the polarization power spectrum 
we will use the integral solution
of the Boltzmann equation \cite{sz}. In the case of scalar 
perturbations for any given Fourier mode $\bi{k}$ 
only $Q^{(S)}$ is generated
in the frame where $\bi{k} \parallel \hat{\bi{z}}$ \cite{kaiser}, 
\begin{eqnarray}
Q^{(S)}(\hat{\bi{n}},\bi{k}) &=& {3 \over 4}(1-\mu^2)\int d\tau
e^{ix\mu} g(\tau)\Pi(k,\tau)
\label{integsol}
\end{eqnarray}
where $x=k (\tau_0 - \tau)$ and $\tau$ is the conformal time 
with $\tau_0$ its present value. Directions in the sky are
denoted with polar coordinates ($\theta$, $\phi$) and 
$\mu=\cos(\theta)$.
We introduced the 
visibility function $g(\tau)=\dot\kappa e^{-\kappa}$, where
$\dot{\kappa}$ is the differential optical depth for Thomson scattering, 
$\dot{\kappa}=an_ex_e\sigma_T$, $a(\tau)$
is the expansion factor normalized
to unity today, $n_e$ is the electron density, $x_e$ is the ionization
fraction and $\sigma_T$ is the Thomson cross section.
The source term $\Pi=\Delta_{T2}^{(S)}
+\Delta_{P2}^{(S)}+
\Delta_{P0}^{(S)}$ was expressed in terms
of temperature quadrupole $\Delta_{T2}^{(S)}$, 
polarization monopole $\Delta_{P0}^{(S)}$ and its quadrupole 
$\Delta_{P2}^{(S)}$. 
Because $U^{(S)}=0$ and $Q^{(S)}$
is only a function of $\mu$
in the  $\bi{k} \parallel \hat{\bi{z}}$
frame it follows $\baredth^2(Q+iU)=\edth^2(Q-iU)$ and so
$a^{(S)}_{2,lm}=a^{(S)}_{-2,lm}$. It is convenient to
introduce two orthogonal combinations $a_{E,lm}=-(
a_{2,lm}+a_{-2,lm})/2$ and $a_{B,lm}=(
a_{2,lm}-a_{-2,lm})/2$. Here $E$ and $B$ 
refer to electric and magnetic type parities \cite{newman} and we have 
chosen the overall sign to agree with the small scale expressions in 
\cite{uros}. Note that our $E$ and $B$ are proportional to $G$ and $C$ in 
\cite{kks}.
We find that $a_{B,lm}^{(S)}=0$ and only $a_{E,lm}^{(S)}$ is non-zero.
The polarization power spectrum is defined as the rotationally invariant 
quantity $C_{l}={1 \over 2l+1}\sum_ma_{lm}^{*}a_{lm}$. 
For $E$ its ensemble average can be obtained by acting twice with spin raising (or
lowering) operator
on equation \ref{integsol} leading to (see \cite{zs} for details)
\begin{eqnarray} 
C_{El}^{(S)}
&=&(3\pi)^2{(l+2)! \over (l-2)!}\int k^2dkP_\phi(k)\nonumber \\
&\times&\left( \int d\tau g(\tau)\Pi(k,\tau){j_l(x) \over x^2}\right)^2,
\end{eqnarray}
where $j_l(x)$ is the spherical Bessel function of order $l$ 
and $P_\phi(k)$ is the primordial power spectrum of scalar 
metric perturbations, usually assumed to be a power law 
$P_\phi(k) \propto k^{n-4}$. 
%Cross-correlation power spectrum 
%between polarization and temperature, $C_{Cl}^{(S)}={1 \over 2l+1}\sum_m
%a_{lm}^{E*}a_{lm}^T$ is given by
%\begin{equation}
%C_{Cl}^{(S)}
%=(4\pi)^2\left({[l+2]! \over [l-2]!}\right)^{1/2}\int k^2dkP_\phi(k)\left(
%\int d\tau g\Pi(k,\tau){j_l(x) \over x^2}\right)^2,
%\end{equation}
 
In the case of tensors the form for $Q$ and $U$ in the frame 
where $\bi{k} \parallel \hat{\bi{z}}$ is \cite{tenspol}
\begin{eqnarray}
Q(\hat{\bi{n}},\bi{k}) &=& -(1+\mu^2)e^{2i\phi}\int d\tau
e^{ix\mu} g\Psi(k,\tau) \; +\;  c.c.\nonumber \\
U(\hat{\bi{n}},\bi{k}) &=& -2i\mu e^{2i\phi}\int d\tau
e^{ix\mu} g\Psi(k,\tau) \; +\;  c.c.,
\end{eqnarray}
where the source is a complex sum over the two independent 
tensor polarization states, 
$\Psi=(\Psi^+-i\Psi^{\times})/2$, and can be expressed in terms of temperature
and polarization multipoles as \cite{tenspol}
$\Psi= 
{1\over10}\tilde{\Delta}_{T0}^{(T)}
+{1\over 7}
\tilde {\Delta}_{T2}^{(T)}+ {3\over70}
\tilde{\Delta}_{T4}^{(T)}
 -{3\over 5}\tilde{\Delta}_{P0}^{(T)}
+{6\over 7}\tilde{\Delta}_{P2}^{(T)}
-{3\over 70}
\tilde{\Delta}_{P4}^{(T)}$. 
This time $\baredth^2(Q+iU)$ and $\edth^2(Q-iU)$ are not equal,
so both $a_{E,lm}^{(T)}$ and $a_{B,lm}^{(T)}$ will be nonzero.
Using a similar procedure as above we obtain
their power spectra 
\cite{zs}
\begin{eqnarray} 
&C&_{El}^{(T)}=
(4\pi)^2\int k^2dkP_h(k)\nonumber \\
&\times&\Big| \int d\tau
g(\tau)\Psi(k,\tau)\Big[-j_l(x)+j_l''(x)+{2j_l(x) \over x^2}
+{4j_l'(x) \over x}\Big]\Big|^2 \nonumber \\
&C&_{Bl}^{(T)}=
(4\pi)^2\int k^2dkP_h(k)\nonumber \\
&\times&\Big| \int d\tau
g(\tau)\Psi(k,\tau)\Big[2j_l'(x)
+{4j_l \over x}\Big]\Big|^2, 
\end{eqnarray}
where $P_h(k)\propto k^{n_T-3}$ is the primordial power spectrum of gravity waves. 
In the small scale limit 
these expressions agree with those derived
previously \cite{uros,tenspol}.

\begin{figure}[t]
\centerline{\psfig{figure=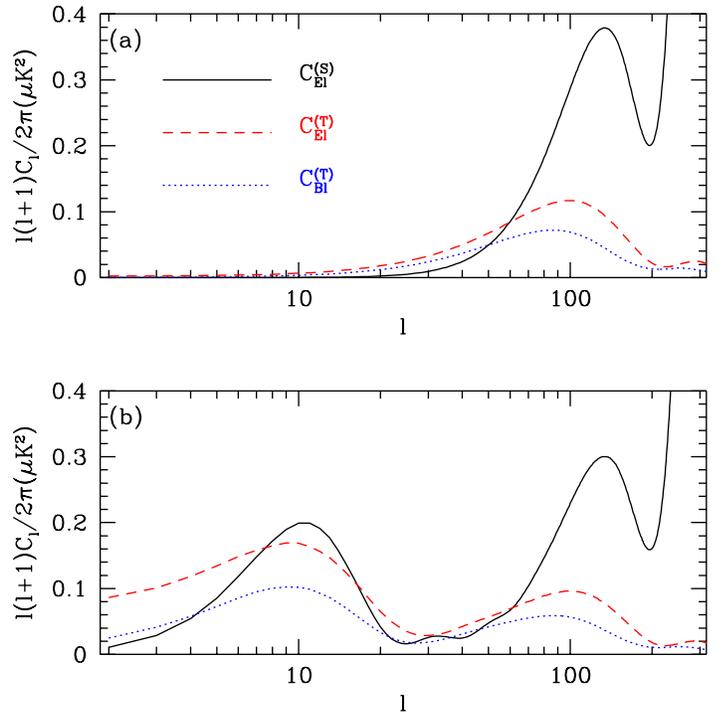,height=4in}}
\caption{Multipole moments for the three polarization spectra for 
no-reionization case (a) and reionized case with optical depth of 0.2
(b).
The underlying model is ``standard CDM'' with $T/S=1$.}
\label{fig1}
\end{figure}

Using the above expressions we may numerically evaluate the power spectra
in various theoretical models. 
We use $T/S$ as the parameter
determining the amplitude of tensor polarization.
Fig.~\ref{fig1} shows the predictions for scalar and tensor 
contribution in standard CDM model 
with no reionization 
(a) and in reionized 
universe with optical depth of $\kappa=0.2$ (b). The latter value is
typical in standard cosmological models \cite{haiman}.
We assumed $T/S=1$ and $n_T=(n_s-1)=-0.15$. In the no-reionization case
both tensor spectra 
peak around $l \sim 100$ and give comparable contributions, although the 
$B$ channel is somewhat smaller. Comparing the scalar and tensor $E$
channels one can see that
scalar polarization dominates for $T/S\lsim 1$. 
Even though 
tensor contribution is larger than scalar at low $l$, the overall power 
there is too small to be measurable. 
Tensor reconstruction in the $E$ channel suffers from similar drawbacks 
as in the case of temperature 
anisotropies: because of large scalar 
contribution cosmic variance prevents one
to isolate very small tensor contributions
\cite{knox}. 
The situation improves if the
epoch of reionization occured sufficiently early that a moderate
optical depth to Thomson scattering is accumulated (Fig.~\ref{fig1}b). 
In this case there is an 
additional peak at low $l$ \cite{zal} and the 
relative contribution of tensor to
scalar polarization in $E$ channel around $l=10$ 
is higher than around $l=100$. Still, if $T/S \ll 1$ 
cosmic variance again 
limits one to extract unambiguously the tensor contribution.
It is in this limit that the importance of $B$ channel becomes crucial. 
This channel is not contaminated by scalar contribution and is only 
limited by noise, so in principle with sufficient noise sensitivity 
one can detect even very small tensor to scalar ratios. Moreover, a detection
of signal in this channel would be a model independent detection of
non-scalar perturbations. In the following we will discuss sensitivity 
to gravity waves using both only $B$ channel information and all 
available information.

We can obtain an 
estimate of how well can tensor parameters be reconstructed by 
using only the $B$ channel and assuming that the rest of cosmological
parameters will be accurately determined from the temperature and $E$
polarization measurements. While this test might not be the most 
powerful it is the least model dependent: any detection in $B$ 
channel 
would imply a presence of non-scalar fluctuations and therefore
give a significant constraint on cosmological models.
Because the $B$ channel does not
cross-correlate with either $T$ or $E$ \cite{uros,kks,zs} 
only its auto-correlation
needs to be considered. 
A useful method to estimate parameter sensitivity for a given experiment 
is to use the Fisher information matrix \cite{jungman,uros,kks,zs}
\begin{eqnarray}
\alpha_{ij}&=&\sum_{l=2}^{l_{\rm max}}{(2l+1)f_{\rm sky} \over 2}
\nonumber \\
&\times&\Big[ C_{Bl}+{4\pi\sigma^2 \over N}e^{l(l+1)\sigma_b^2}\Big]^{-2}
\left( {\partial C_{Bl} \over \partial s_i} \right)\left(
{\partial C_{Bl} \over \partial s_j} \right),
\end{eqnarray}
where $f_{\rm sky}$ is the sky coverage.
Receiver noise
can be parametrized by $4 \pi \sigma^2/N$, where $\sigma$ is the noise 
per pixel and $N$ is the number of pixels. Typical values are
$(0.15 \mu{\rm K})^2$ 
for MAP
and $(0.025\mu{\rm K})^2$ for the most
sensitive Planck bolometer channel
in one year of observation.
In our case the parameters $s_i$ can be $T/S$ and $n_T$, so that the matrix is
only 2x2. The error on each parameter 
is given by
$(\alpha^{-1}_{ii})^{1/2}$ 
if the other parameter is assumed to be unknown 
and $(\alpha_{ii})^{-1/2}$ if the other parameter is assumed to be known.
Using this expression we may calculate the experiment
sensitivity to these parameters. 
Current inflationary models and 
limits from large scale structure and COBE 
predict $T/S$ to be less than unity. 
Figure \ref{fig1} shows that the expected amplitude in this case 
is below 0.5$\mu$K.
We find that MAP is not sufficiently sensitive in $B$ channel 
to detect these low levels.
On the other hand, Planck will be much more
sensitive and can detect $T/S >0.3$ if tensor index $n_T$ is assumed
to be known (for example through the
consistency relation). For the underlying model with 
$T/S=1$ one can determine it
with an error $\Delta( T/S) \sim 0.1$. If tensor index is not known then 
a combination of the two parameters, which corresponds to the total
power under the $B$ curve in figure \ref{fig1}, can still be 
determined with the same accuracy. 

Separate determination of the tensor amplitude and slope
from the $B$ channel is only possible 
in reionized models. 
In the no-reionization model the contribution to $B$ is very
narrow in $l$ space and the leverage on $n_T$ independent of $T/S$ is small,
so that 
the correlation coefficient $\alpha_{12}/(\alpha_{11}\alpha_{22})^{1/2}$
is almost always close to unity.   
A modest amount of reionization improves the separation; 
in the reionized models the power spectrum for $B$ is bimodal 
(figure \ref{fig1})
and the overall signal is higher, which gives
a better leverage on $n_T$ independent of $T/S$.
For $\kappa=0.2$ the Planck errors are $\Delta (T/S) \sim 0.15$ and
$\Delta n_T \sim 0.1$ for the underlying model with $T/S = 1$. 
These results depend on
the overall amplitude
relative to the noise level. As long as both peaks can be separated 
from the noise one can determine the tensor slope, which allows to 
test the inflationary consistency relation.

Combining all the information by adding temperature, $E$ polarization 
and their cross-correlation
further improves these estimates. In this case other parameters
that affect scalar modes 
such as baryon density, Hubble constant or cosmological constant
enter as well and the results become more
model dependent \cite{zss}. 
Fisher information matrix has to be generalized to 
include all the parameters that can be degenerate with the tensor 
parameters. 
The results 
depend on the class of models and number of parameters one restricts
to in the analysis, as opposed to the results based on $B$ channel above,
which  depend only on the two main parameters that characterize the
gravity wave production. 
As a typical example, for $T/S=0.1$ and $\kappa=0.1$
one can determine $\Delta (T/S)=0.05$ and $\Delta n_T=0.2$ with 
Planck \cite{zss}. 
These errors improve 
further if a model with higher $T/S$ or 
$\kappa$ is assumed. 
For the same underlying 
model without using polarization the expected errors are 
$\Delta T/S \sim 0.26$ and $\Delta n_T\sim 1$, significantly worse
than with polarization. 
Even for MAP the limits on $T/S$ improve by a factor of 2 when 
polarization information is included.

To summarize the above discussion, 
future CMB missions are likely to reach the sensitivities needed to measure
(or reject) a significant production of primordial gravity waves in 
the early universe through polarization measurements, which will vastly 
improve the limits from temperature measurements only and allow 
a test of
consistency relation.
The more challenging question is 
the foreground subtraction at the required level. 
At low frequencies radio point sources and synchrotron emission 
from our galaxy dominate the foregrounds
and both are polarized at a 10\% level.  
Their contribution 
decreases at higher frequencies and with several frequency 
measurements one can subtract these foregrounds at 
frequencies around 100 GHz
at the required microkelvin level. 
At even higher frequencies dust 
is the dominant foreground, but
is measured to be only a few percent polarized \cite{dust}. 
We hope that the signature
of gravity waves discussed here 
would provide further motivation to pursue the feasibility studies of 
polarization measurements.

While we only discussed scalar and tensor modes, vector modes, if present
before recombination, will also contribute to both polarization 
channels and so could contaminate the signature of gravity waves.
At present there are no viable cosmological models that would
produce a significant contribution of vector modes without a comparable
amount of tensor modes. In inflationary models vector modes, even if
produced during inflation, decay away and are not significant during 
recombination. In topological 
defect models nonlinear sources continuously 
create both vector and tensor modes and so some of the signal in $B$ 
channel could be caused by vector modes. Even in these models however 
some fraction of signal in $B$ will still be generated
by tensor modes and in any case, absence of signal in $B$ channel would
rule out such models. Polarization thus offers a unique way to probe 
cosmological models that is within reach of the next generation of CMB 
experiments.

\vfil\eject


\begin{references}
\bibitem{circular} Circular polarization cannot be
generated in the early universe through the process of Thomson
scattering, so Stokes parameter $V$ is 0.
\end{references}
\end{document}